\documentclass{IEEEtran}
\usepackage{cite}
\usepackage{amsmath,amssymb,amsfonts}
\usepackage{algorithmic}
\usepackage{graphicx}
\usepackage{textcomp}
\usepackage{xcolor}		

\usepackage{color}

\def\BibTeX{{\rm B\kern-.05em{\sc i\kern-.025em b}\kern-.08em
    T\kern-.1667em\lower.7ex\hbox{E}\kern-.125emX}}
\graphicspath{{./figures/}}
\begin{document}
\title{High-Speed PAM4-Based Optical SDM Interconnects With Directly Modulated Long-Wavelength VCSEL}

\author{Joris Van Kerrebrouck, \IEEEmembership{Student Member, IEEE},
Xiaodan Pang, \IEEEmembership{Member, IEEE},
Oskars Ozolins, \IEEEmembership{Member, IEEE},
Rui Lin,
Aleksejs Udalcovs, \IEEEmembership{Member, IEEE},
Lu Zhang,
Haolin Li, \IEEEmembership{Student Member, IEEE},
Silvia Spiga,
Markus-Christian Amann, \IEEEmembership{Fellow, IEEE},
Lin Gan,
Ming Tang, \IEEEmembership{Senior Member, IEEE},
Songnian Fu, Richard Schatz, Gunnar Jacobsen, Sergei Popov, Deming Liu, Weijun Tong,
Guy Torfs, \IEEEmembership{Member, IEEE},
Johan Bauwelinck, \IEEEmembership{Senior Member, IEEE},
Jiajia Chen, \IEEEmembership{Senior Member, IEEE},
Xin Yin, \IEEEmembership{Member, IEEE}

\thanks{Manuscript received xx xx, 2018; (Joris Van Kerrebrouck and Xiaodan Pang contributed equally to this work, corresponding author: Jiajia Chen)}
\thanks{J. Van Kerrebrouck, H. Li, G. Torfs, J. Bauwelinck, and X. Yin are with Ghent University - imec, IDLab, Department of Information Technology, Gent 9000, Belgium (e-mails: joris.vankerrebrouck@ugent.be; haolin.li@UGent.be; guy.torfs@ugent.be; johan.bauwelinck@ugent.be; xin.yin@ugent.be). }
\thanks{X. Pang, R. Lin, L. Zhang, R. Schatz, S. Popov are with KTH Royal Institute of Technology, Kista 164 40, Sweden (e-mails:
xiaodan@kth.se; rulin@kth.se; lu8@kth.se; rschatz@kth.se; sergeip@kth.se).}
\thanks{L. Zhang is with the State Key Laboratory of Advanced Optical Communication System and Networks, Shanghai Jiao Tong University, Shanghai 200240, China (email: luzhang\_sjtu@sjtu.edu.cn).}
\thanks{J. Chen is with MOE International Laboratory for Optical Information Technologies, South China Academy of Advanced Optoelectronics, South China Normal University, Guangzhou 510006, China, and KTH Royal Institute of Technology, Kista 164 40, Sweden (e-mail: jiajiac@kth.se).}
\thanks{R. Lin, L. Gan, M. Tang, S. Fu and D. Liu are with the Wuhan National lab for Optoelectronics (WNLO), Huazhong University of Sci\&Tech (HUST), Wuhan 430074, China (e-mails: lingan@hust.edu.cn;   tangming@hust.edu.cn; songnian@mail.hust.edu.cn; dmliu@hust.edu.cn).}
\thanks{S. Spiga and M. C. Amann are with the Walter Schottky Institut, Technische Universit\"at M\"unchen Garching 85748, Germany (e-mails: silvia.spiga@wsi.tum.de; amann@wsi.tum.de).}
\thanks{W. Tong is with the Yangtze Optical Fibre and Cable Company Ltd (YOFC), Wuhan 430073, China (e-mail: tongweijun@yofc.com).}
}

\maketitle

\begin{abstract}
This paper reports the demonstration of high-speed PAM-4 transmission using a 1.5-$\mu$m single-mode vertical cavity surface emitting laser (SM-VCSEL) over multicore fiber with 7 cores over different distances. We have successfully generated up to 70 Gbaud 4-level pulse amplitude modulation (PAM-4) signals with a VCSEL in optical back-to-back, and transmitted 50 Gbaud PAM-4 signals over both 1-km dispersion-uncompensated and 10-km dispersion-compensated in each core, enabling a total data throughput of 700 Gbps over the 7-core fiber. Moreover, 56 Gbaud PAM-4 over 1-km has also been shown, whereby unfortunately not all cores provide the required 3.8 $\times$ 10$^{-3}$ bit error rate (BER) for the 7\% overhead-hard decision forward error correction (7\% OH-HDFEC). The limited bandwidth of the VCSEL and the adverse chromatic dispersion of the fiber are suppressed with pre-equalization based on accurate end-to-end channel characterizations. With a digital post-equalization, BER performance below the 7\% OH-HDFEC limit is achieved over all cores. The demonstrated results show a great potential to realize high-capacity and compact short-reach optical interconnects for data centers.
\end{abstract}


\section{Introduction}
\label{sec:introduction}
\IEEEPARstart{R}{ecent} forecasts predict a compound annual growth rate of 25\% of the total data center traffic for the years from 2016 to 2021 \cite{Cisco}. Among this, the traffic within data centers rises more rapidly with 31.9\% per year. The total computing power of the top 500 high-performance computers grow annually 60\%. The use of artificial intelligence clusters with tensor processing units running data-hungry algorithms, raises the bar even higher for data-center interconnections. In addition to the growing number of servers in hyper-scale data centers \cite{datacenter}, the number of interconnections grows even faster due to newly evolved architectures and topologies such as spline-leaf and east-west configurations. On the other hand, the cost, reliability, power consumption and footprint have to be optimized simultaneously with the increasing data throughput.

To cope with the limited bandwidth of optical devices such as VCSELs and photodiodes, and the influence of chromatic dispersion, PAM-4 is more advantageous than non-return-to-zero (NRZ). More advanced modulation formats, like discrete-multi-tone (DMT) and the multiband approach of carrierless amplitude phase (MultiCAP) modulation, support high-speed single-lane transmissions with even higher spectral efficiency than PAM-4, but with an increased complexity as a consequence. Therefore, parallel channels can be exploited as an alternative technique to increase the data throughput. In this paper, the spatial division multiplexing (SDM) with a multi-core fiber (MCF) is chosen as one of the potential solutions. Reducing cost and power consumption simultaneously is a tremendous challenge \cite{Kuchta}. SDM can benefit from lower cost and power consumption thanks to the integration of many system components while also providing a very high throughput owing to the parallel channels \cite{Richardson}. A VCSEL array can seamlessly extend the potential of SDM to a relatively longer-term evolution towards “parallelism”, helping scale up the lane count per fiber and reduce cost per bit, therefore resulting in an improved energy efficiency. There are techniques to butt couple a VCSEL array directly to a MCF facet \cite{Karppinen}, making a complex MCF lantern fan-out obsolete. Research has been performed on MCF connectors to enable practical implementations \cite{Watanabe}. A few recent works \cite{Kerrebrouck1} \cite{Eiselt} and \cite{Xie} on high data rate transmission with 1.5-$\mu$m SM-VCSELs have also shown promising results towards the combination of a VCSEL with a MCF for different modulation formats.

This paper is an invited extension of our work presented at OFC 2018 \cite{Pang}. In Section II, we will explain the performance of the major building blocks, namely the SM-VCSEL and the MCF. In Section III, the results of 50, 56, 64 and 70~Gbaud multilane PAM-4 over an optical back-to-back (B2B) link will be discussed. 50 Gbaud PAM-4 over 1-km is achieved with all fiber cores below 7\% OH-HDFEC, and even 56~Gbaud PAM-4 over 1-km is achievable for some cores with $<$\,7\% OH-HDFEC. Moreover, over 10-km fiber distance with a dispersion compensation modules (DCM), 50~Gbaud PAM-4 is demonstrated for all cores with $<$\,7\% OH-HDFEC. Finally, Section IV concludes this paper.

\section{DEVICES CHARACTERIZATION AND EXPERIMENTAL SETUP}

\subsection{InP-based Long-Wavelength VCSEL}
\begin{figure}[!t]
\centerline{\includegraphics[width=0.9\columnwidth]{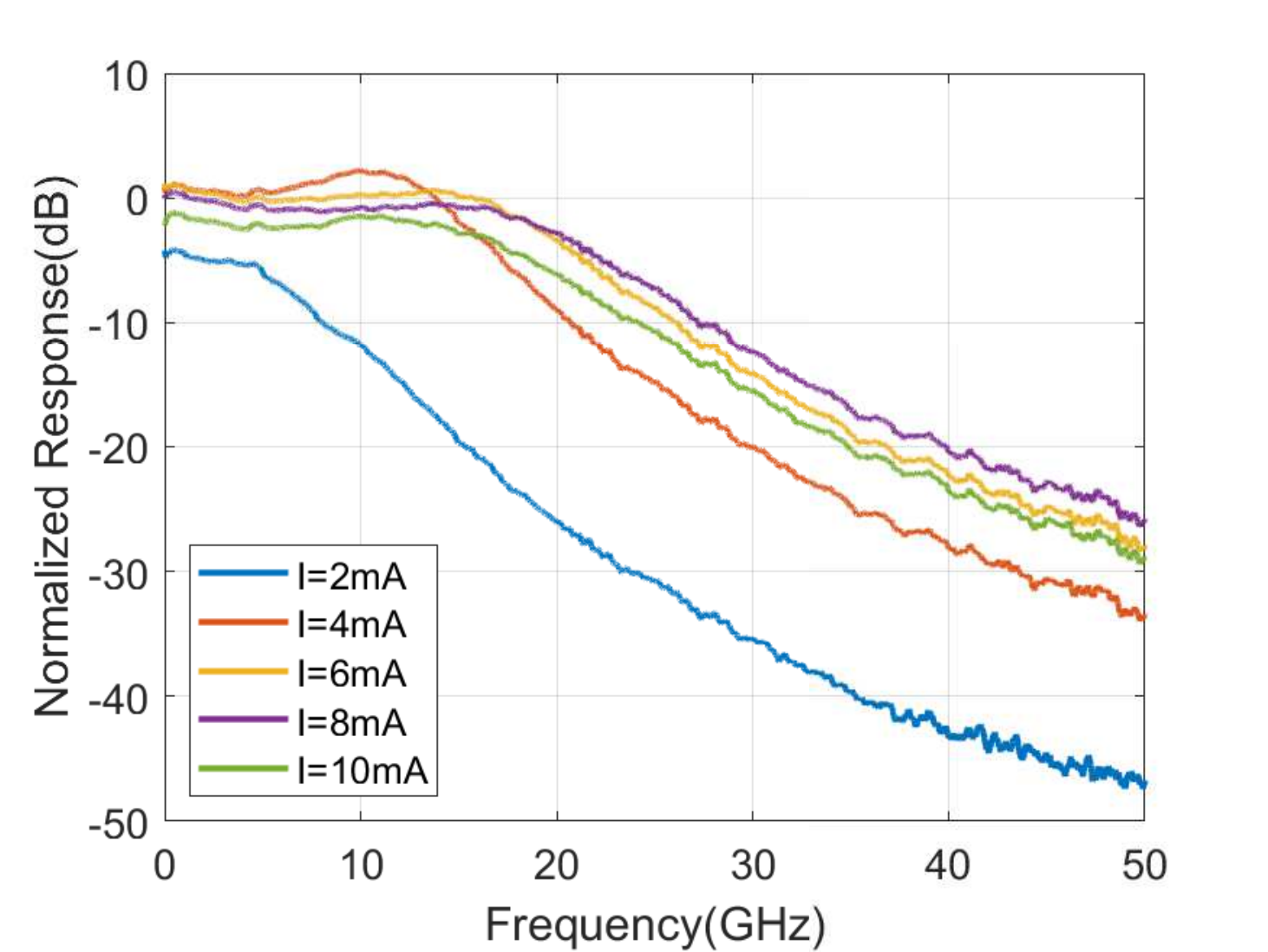}}
\caption{Small-signal S21 responses of the 1550~nm VCSEL, normalized to the S21 of the VCSEL operating at a bias of 8~mA. }
\label{fig1}
\end{figure}

Compared to GaAs-based VCSELs, InP-based VCSELs emit at longer wavelengths i.e. 1.3 and 1.55-$\mu$m. These longer wavelength VCSELs have a lower power consumption (due to the lower band gap) and these wavelengths have ten times lower loss in silica-based optical fibers. The performances of the 1.5$\mu$-$\lambda$ cavity VCSEL used in this paper were first presented in \cite{Spiga1} and \cite{Spiga2}. The frequency response of the VCSEL is shown in Fig. \ref{fig1}. The 3~dB small-signal modulation bandwidth of 20~GHz is obtained at a bias current of 7~mA at room temperature. The approach used to achieve this high intrinsic small-signal modulation bandwidth is the reduction of the laser effective cavity length. By means of dielectric distributed Bragg reflectors (DBRs), high refractive index contrast between the lambda-quarter layers is achieved, leading to short penetration depth of the field in the DBRs and reducing the overall cavity length. To reach a high modulation frequency, a high relaxation frequency is needed which is inversely proportional to the square root of the photon lifetime. By reducing the cavity length, the photon lifetime is minimized. Other modulation frequency restrictions are coming from parasitic effect and thermal issues, and these are overcome by an involved doping profile. The optical gain is provided by an active region with seven AlGaInAs quantum wells and current confinement is achieved by a p+ –AlGaInAs/n+ –GaInAs buried tunnel junction (BTJ) of 4-$\mu$m. 

\begin{figure}[!t]
\centerline{\includegraphics[width=\columnwidth]{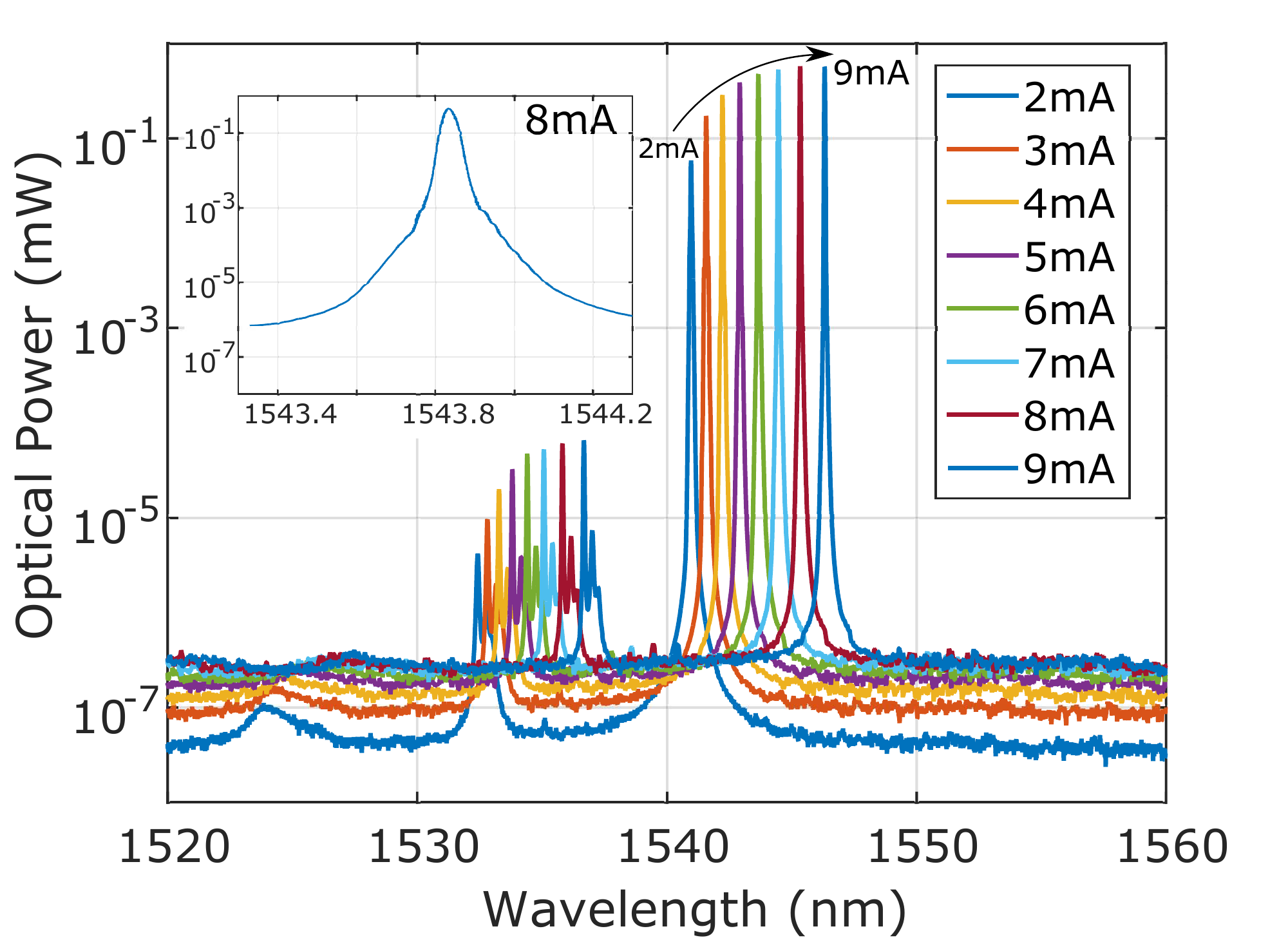}}
\caption{Optical spectrum of the VCSEL measured for different bias currents, going from 2~mA to 9~mA. The enlarged optical spectrum around the fundamental wavelength of the VCSEL is measured at a bias current of 8~mA with a resolution bandwidth of 30~pm.}
\label{fig2}
\end{figure}

\begin{figure}[!t]
\centerline{\includegraphics[width=\columnwidth]{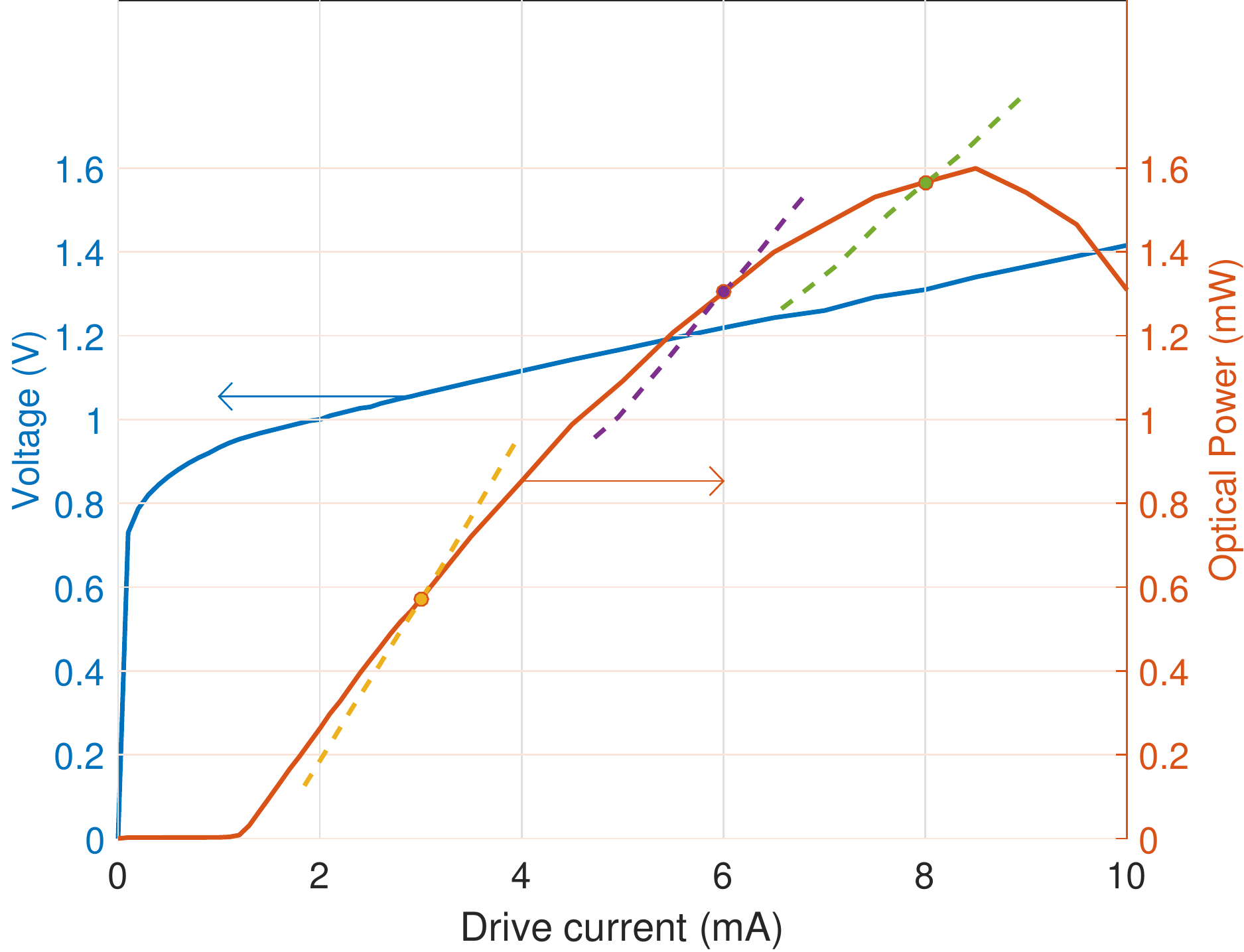}}
\caption{The L-I-V characteristic of the 1550~nm VCSEL. The left axis is the voltage across the VCSEL at a given bias current. Right axis is the static optical output power in function of the bias current. The dashed lines represent the dynamic optical output power measured wit a modulated RF signal when the VCSEL is biased at 3mA, 6mA and 8mA, respectively.}
\label{fig3}
\end{figure}

In Fig. \ref{fig2} lower optical modes are observed for this VCSEL. A side-mode suppression ratio of 40~dB is achieved, which is enough to enable the VCSEL's single-mode operation.
The voltage and output power versus the bias current are presented in the light-current-voltage (L-I-V) characteristic curves of the VCSEL in Fig. \ref{fig3}. The threshold current is approximately 1.5~mA and an optimal current to maximize the optical power is found to be around 8 mA, also called rollover point, hereafter the VCSEL output power starts to degrade due to the thermal effect. However, the dynamically modulated light does not saturate due to the thermal effect. Because the temperature remains constant at the temperature set by the bias current and the rapid changes of current caused by modulating the VCSEL are averaged out. The dynamic output power behavior is measured with a triangle waveform with a frequency of 10~MHz and 130~mVpp voltage swing while the bias current is recorded at 3~mA, 6~mA and 8~mA. The results are represented by the dashed lines in Fig. \ref{fig3}. The measurements show that the fast modulated optical signal is not bounded by the thermal effect.

\begin{figure}[!t]
\centerline{\includegraphics[width=\columnwidth]{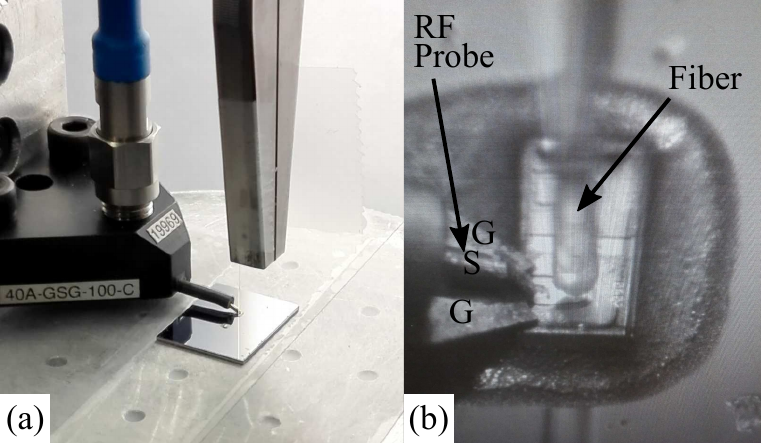}}
\caption{(a) The experimental VCSEL-probing setup. (b) Microscope picture with the GSG RF probe and the lensed fiber to capture the light. }
\label{fig4}
\end{figure}

Fig. \ref{fig4}(a) shows the  experimental VCSEL-probing setup where the VCSEL die is fixed on a silicon plate and is held in place by a vacuum chuck. A 40 GHz GSG RF probe with a probe pitch of 100 $\mu$m from Picoprobe is used. Fig. \ref{fig4}(b) shows a microscope picture of the probing station where the ground-signal-ground (GSG) needles with a pitch of 100-$\mu$m feed the RF signal to the VCSEL, and the light is captured with a lensed fiber.

\subsection{Multi-Core Fiber}
\begin{figure}[!t]
\centerline{\includegraphics[width=\columnwidth]{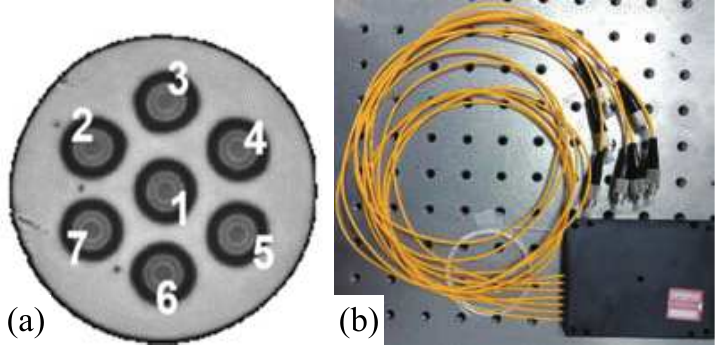}}
\caption{(a) Cross-section of the YOFC MCF-7-42/150/250. (b) The multi-core fiber fan-in or fan-out module FAN-7-42.}
\label{fig5}
\end{figure}

In this paper, we employed a 1-km 7-core MCF and a 10-km 7-core MCF \cite{YOFC}. The cross-section of the 7-core MCF is depicted in Fig. \ref{fig5}(a). The stack and draw process is applied to fabricate these homogeneous MCFs, and therefore, the properties of different cores are almost identical. The MCF is designed for short-reach communications with ultra-low crosstalk \cite{Li} and \cite{Tu}. The cladding diameter of the MCF is 150-$\mu$m and the average core pitch is 41.5-$\mu$m. The crosstalk between adjacent cores is suppressed to be as low as -45dB/100km. The attenuation is less than 0.2~dB/km in C-band and the chromatic dispersion of the MCF is 17.1~ps/nm/km. 
Low-loss and reliable connectivity between the MCF and the SSMFs are ensured for the MCF based optical links by using fan-in/fan-out modules. The fabricated fan-in/fan-out module \cite{YOFC} is shown in Fig. \ref{fig5}(b). The crosstalk between adjacent cores in the fan-in/fan-out modules is measured to be less than -50 dB, and the insertion loss per fan-in/fan-out is less than 1.5 dB.

\subsection{Experimental Setup}
\begin{figure}[!t]
\centerline{\includegraphics[width=\columnwidth]{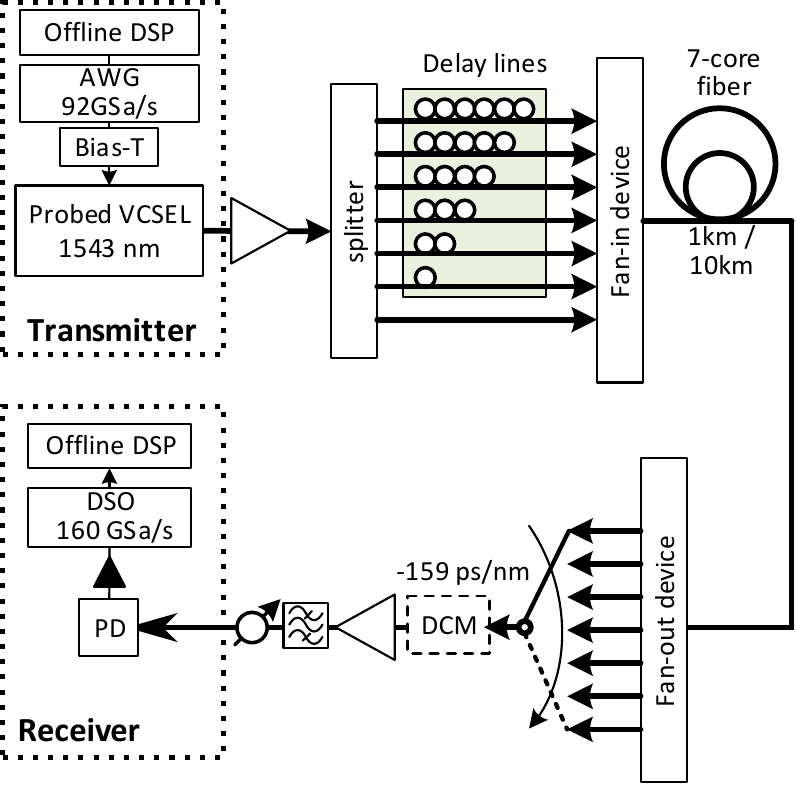}}
\caption{Experimental setup.}
\label{fig6}
\end{figure}

In Fig. \ref{fig6}, the block diagram of our experimental setup is shown. An arbitrary waveform generator (AWG) Keysight M8196A, with 92 GSA/s, 32 GHz analog bandwidth and 8-bit vertical resolution, converts the pre-equalized PAM-4 signal to an analog signal. The PAM-4 symbols are generated offline from a pseudo random bit sequence with a word length $2^{15}-1$ (PRBS-15), a raised cosine filter with a 0.15 roll-off factor is used as pulse shape. Frequency-domain channel pre-equalization is performed on the PAM-4 signal based on the characterized end-to-end channel frequency response. After pre-equalization, the signal is loaded to the AWG for modulating the VCSEL. The amplitude of the driving signal is set to be 700~mVpp. The VCSEL is operated at room temperature without active cooling, and the optimal driving current is found to be 7.8~mA, considering both bandwidth and output power.

The VCSEL output is amplified and split into seven lanes with proper delays in order to decorrelate the lanes from each other, emulating parallel independent channels entering the fan-in device for the 7-core fiber. After transmission over the 1-km or 10-km MCF link, a fan-out device is used to couple the signals to single-core fibers, prior to the receiver. For 1-km MCF transmission the link is uncompensated while a fixed dispersion compensation module (DCM) of -159~ps/nm is used for the 10-km MCF case yielding a residual dispersion of approximately 12~ps/nm. After transmission over the MCF, the signal is detected by a PIN photodiode (PD) with $>$90~GHz bandwidth. Due to the low PD responsivity and the absence of a transimpedance amplifier (TIA), an automatic gain-controlled erbium-doped fiber amplifier (EDFA) with fixed output power is employed as a pre-amplifier. An optical filter is used for suppressing the amplified spontaneous emission (ASE) that arises from the EDFA. A variable optical attenuator (VOA) regulates the optical power incident on the photodiode. In practical systems, this receiver structure can be substituted with a PIN-TIA. The PD output signal is sampled at 160~GSa/s by a 63~GHz real-time digital storage oscilloscope (DSO) (Keysight DSAZ634A) and processed further offline using Matlab\textregistered. 

\section{RESULTS AND DISCUSSIONS}
\begin{figure}[!t]
\centerline{\includegraphics[width=\columnwidth]{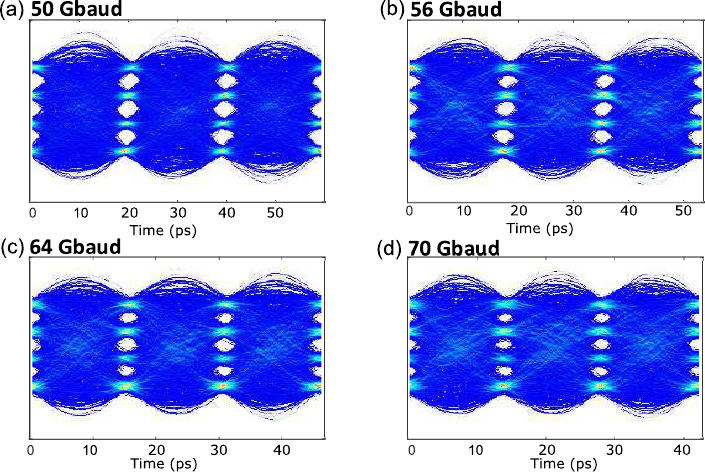}}
\caption{B2B equalized eye diagrams at 7 dBm RoP.}
\label{fig7}
\end{figure}

In this section, the performance of the PAM-4 transmission over the MCF is evaluated. For comparison, we evaluate different symbol rates from 50 to 70~Gbaud. To further explore the impact of fiber length on the performance, the performances for optical B2B, 1-km and 10-km are compared. The first experiment was performed for B2B to obtain the BER limit of the optical devices, serving as the performance reference. The eye diagrams obtained after pre-equalization in the AWG at a received optical power (RoP) of 7~dBm are shown in Fig. \ref{fig7}. Clear eye openings are observed for all different symbol rates considered, despite the slight eye skew caused by signal-dependent rise/fall time during VCSEL modulation \cite{Castro}. 

\begin{figure}[!t]
\centerline{\includegraphics[width=0.9\columnwidth]{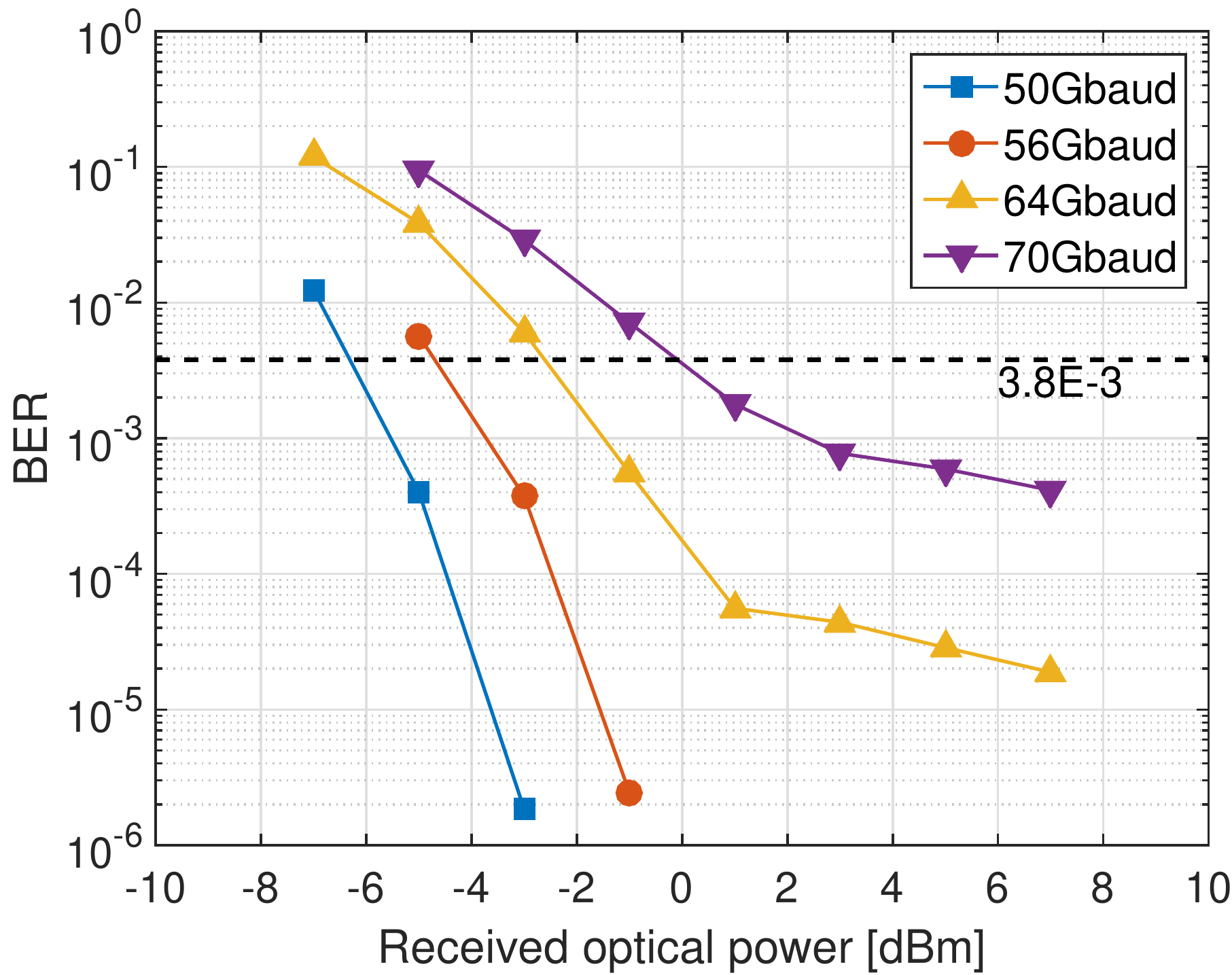}}
\caption{The B2B BER results for 50, 56, 64 and 70 Gbaud PAM-4 with equalization.}
\label{fig8}
\end{figure}

This eye skew may form a limiting factor for an error-free transmission due to the differences in optimum sampling time for the upper, middle, or lower eyes. The eye skew could be reduced by increasing the bias current. However, this is limited by the maximally allowed bias current due to thermal heating. In this paper, the eye skew has been left unresolved, one common sampling time stamp is chosen for all eye decisions. The utilization of multiple timing circuits, each dedicated to sample the top, middle or bottom eyes, increases the complexity and the optical link dependency of the transceivers. Fig. \ref{fig8} shows the BER curves for the modulated PAM-4 signals of different baud rates in an optical B2B link. The 7\% OH-HDFEC limit of 3.8 $\times$ 10$^{-3}$ BER has been successfully achieved at ~0 dBm received optical power (RoP) for up to 70~Gbaud PAM-4 signals.  
while 5-tap feedforward (FF) and 5-tap feedback (FB) equalizers at the receiver were used thanks to the effective channel pre-equalization at the transmitter based on the end-to-end link. For the 50, 56 and 64~Gbaud signals, 3-tap FF + 3-tap FB equalizers are already sufficient to eliminate the residual inter-symbol interference (ISI). The KP4-FEC (2.4 $\times$ 10$^{-4}$) with an overhead of even 3\% is achievable for the 50, 56 and 64 Gbaud at around 1 dBm RoP. 

\subsection{50~Gbaud 1-km}
\begin{figure}[!t]
\centerline{\includegraphics[width=0.8\columnwidth]{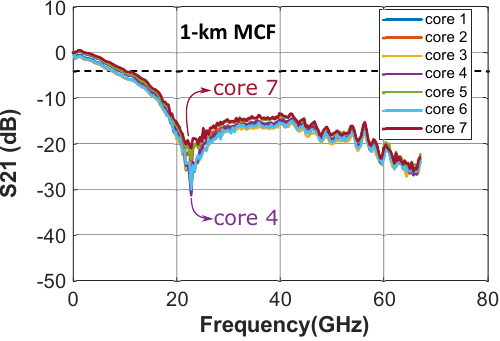}}
\caption{The frequency response (S21) of the total E-O-E link with 1-km MCF.}
\label{fig9}
\end{figure}

With 1-km MCF fiber, the bandwidth limitation is immediately visible in the Fig. \ref{fig9}. The normalized transmission (S21) reveals the effect of chromatic dispersion of the fiber where a notch at 23 GHz is observable. For 1-km fiber used in C-band we expect the first notch at around 60 GHz, however, due to the chirp of the VCSEL, the chromatic dispersion notch in the S21 appears unfavorably at a much lower frequency. This has also been presented in previous experiments where the linewidth enhancement factor ($\alpha_{H}=\frac{2\beta}{A}$) of this VCSEL is found to be 12 approximately \cite{Kerrebrouck2}. To adopt directly-modulated lasers (DML), in particular VCSELs, the linewidth enhancement factor needs to be minimized to achieve a high data rate link or to reach long distance. 

\begin{figure}[!t]
\centerline{\includegraphics[width=\columnwidth]{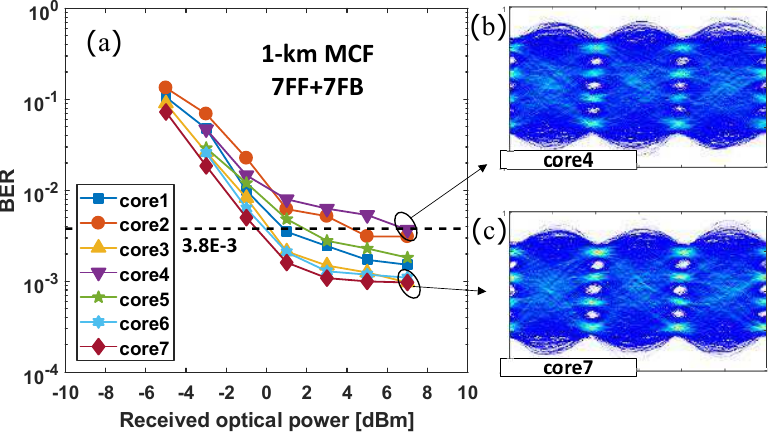}}
\caption{BER results for 50~GBaud PAM-4 signal after 1-km MCF and eye diagrams at 7~dBm RoP for the best and worst cores.}
\label{fig10}
\end{figure}

Fig. \ref{fig10}(a) shows the BER performance after the 1-km MCF transmission with 50~Gbaud PAM-4 signals. For the uncompensated 1-km MCF case, channel pre-equalization is done at the transmitter in combination with receiver side 7-tap FF and 7-tap FB equalizers to successfully recover the received signals in all the cores with a BER performance below the 7\%-OH-HDFEC limit. Selected eye diagrams of the best and worst cores are also shown in Fig. \ref{fig10}(b) and (c).

\begin{figure}[!t]
\centerline{\includegraphics[width=0.8\columnwidth]{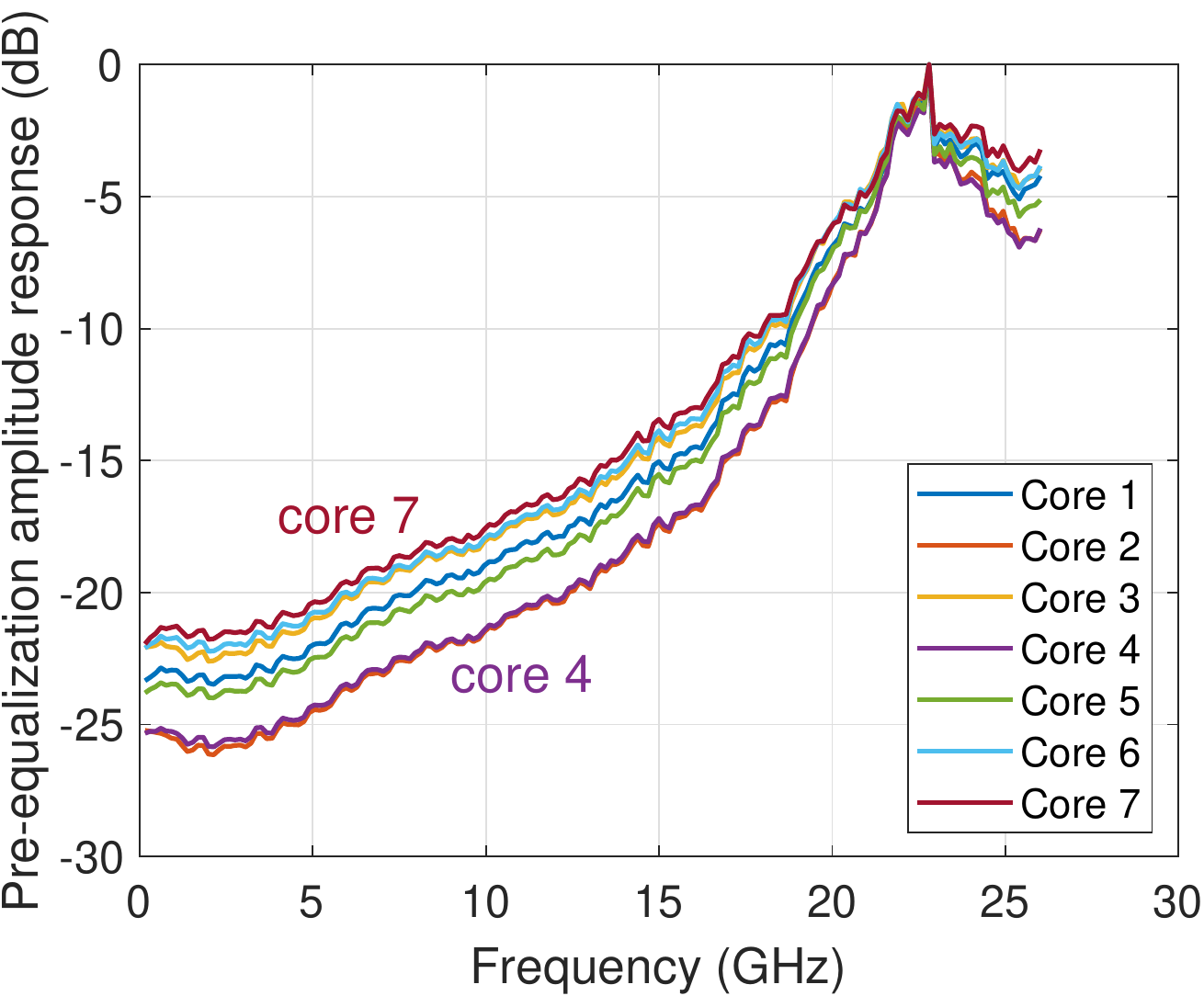}}
\caption{Transmitter pre-equalization channel response for each core of the E-O-E 1-km MCF link.}
\label{fig11}
\end{figure}

The transmitter pre-equalization channel response is shown in Fig. \ref{fig11}. The channel is measured by connecting the Keysight M8196A AWG to the optical link and measuring the analog electrical output with the Keysight real-time oscilloscope DSAZ634A. By using built-in calibration tools of the equipment, the channel is measured and can be used for the zero-forcing pre-equalization. The cut-off frequency of the equalizer is set to 26~GHz. Frequencies lower than the cut-off are attenuated according to the channel response to the cut-off frequency. At the transmitter, the pre-equalizer is applied on the PAM-4 signal together with the pulse shaping filter and then the digitally equalized signal is normalized to utilize the full scale of the DACs in AWG. By limiting a maximum cut-off frequency for the equalizer, the low frequencies are not too much suppressed, and the high frequencies (higher than the Nyquist frequency) are not unnecessarily amplified. 

A zero-forcing equalizer will boost the signal around the notch frequency tremendously, because the algorithm tries to invert the channel response. Note that, this boosting due to transmitter pre-equalization occupies also the dynamic range of the AWG DACs, which may lead to smaller vertical eye openings and hence degraded performances. For example, at core 4, the measured notch is -30~dB @ 23~GHz (Fig. \ref{fig9}). Hereby, the zero-forcing equalizer will extremely boost the signals around the 23~GHz frequency. To fit the equalized signal into the full-scale of the DAC, the low frequencies need to be attenuated (see core 4 in Fig. \ref{fig11}), which degrade the BER performance (see core 4 in Fig. \ref{fig10}). On the other hand, core 7 measured a notch of -22 dB and less equalization is needed, resulting in a bigger eye and a better BER performance.

\begin{figure}[!t]
\centerline{\includegraphics[width=0.9\columnwidth]{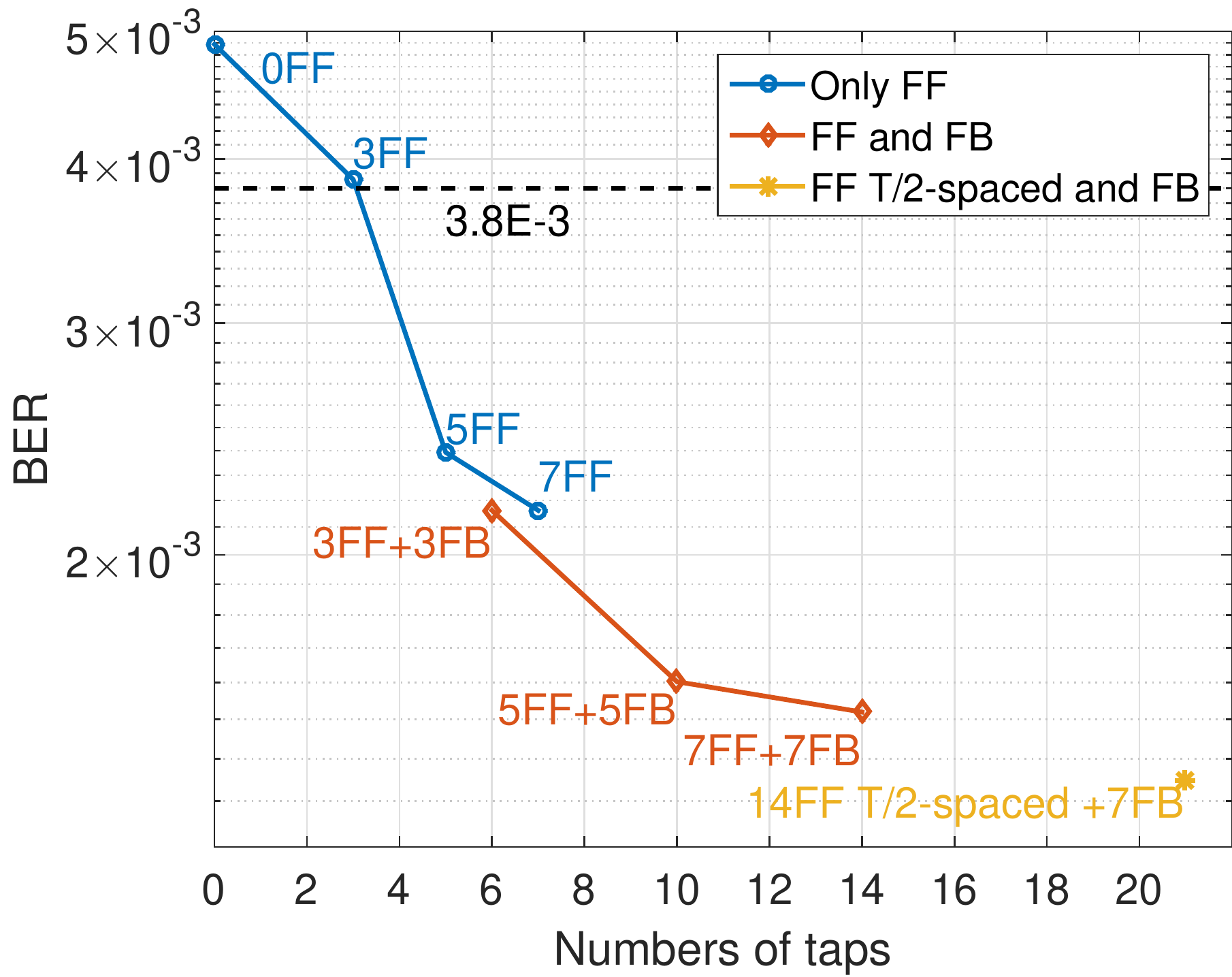}}
\caption{BER for 50~GBaud PAM-4 signal after 1-km MCF and with a received optical power of 7~dBm. The BER is calculated with different equalizer combinations and various number of taps.}
\label{fig12}
\end{figure}

The influence of the receiver equalizer is presented in Fig. \ref{fig12} by giving the BER for the 50~Gbaud PAM-4 over 1-km MCF in core 1 with a received optical power of 7~dBm. Three different equalizer combinations are compared: only feedforward (FF) equalizer, feedforward in combination with feedback (FB) equalizer and a feedforward together with a half-symbol spaced feedback equalizer. The number of taps (the sum of the FF and the FB taps) varied from 0 to 21. The FF equalizer improves the BER from 4.9 $\times$ 10$^{-3}$ without any equalization to a BER of 2.2 $\times$ 10$^{-3}$ with 7 taps FF. The adaptive FF equalizer in the receiver only resolves the remaining linear post- and precursor ISI, originated from the little variations in the different cores. To further reduce the BER, an additional FB equalizer is used together with the FF equalizer. Due to the non-linear behavior of the VCSEL in combination with linearly sensitive PAM4 modulation, the use of an FB is very beneficial. An FF can help further remove the ISI since the FB cannot resolve a pre-cursor phenomenon. The additional FB equalizer provides an improvement in BER down to 1.5 $\times$ 10$^{-3}$. By changing the FF taps spacing, like half-symbol spaced taps, a small improvement can be obtained (1.35 $\times$ 10$^{-3}$ BER), albeit with a doubling of the number of taps. Half-symbol spaced FF equalizer has the practical ability to solve the timing error to some extent, but commonly needs 2x oversampling ADCs \cite{pam4_designcon}.

\subsection{50~Gbaud 10-km}
\begin{figure}[!t]
\centerline{\includegraphics[width=0.8\columnwidth]{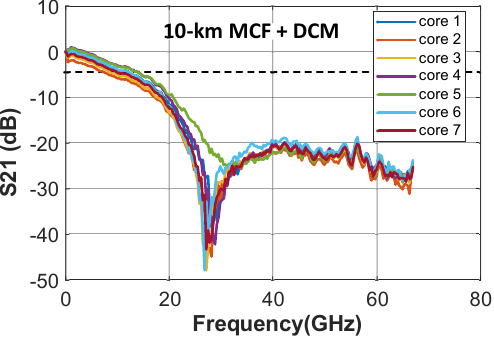}}
\caption{The frequency response of the total E-O-E link with 10-km MCF and DCM.}
\label{fig13}
\end{figure}

In the 10 km MCF case, a dispersion-compensation module for 159~ps/nm is used to compensate the chromatic dispersion of 17.1~ps/(nm km) @ 1550~nm  \cite{YOFC} of 10~km MCF.
The obtained residual chromatic dispersion of 12~ps/nm corresponds to a virtual fiber length of 700~m. The virtual shorter fiber length shifts the chromatic dispersion notch to a higher 27~GHz, providing a slightly higher bandwidth than in the uncompensated 1~km link (chromatic dispersion notch at 23 GHz), as shown in Fig. \ref{fig13}. Due to the 10 times longer fiber lengths and the dispersion-compensation, the dispersion variations between the cores are more pronounced.

\begin{figure}[!t]
\centerline{\includegraphics[width=0.8\columnwidth]{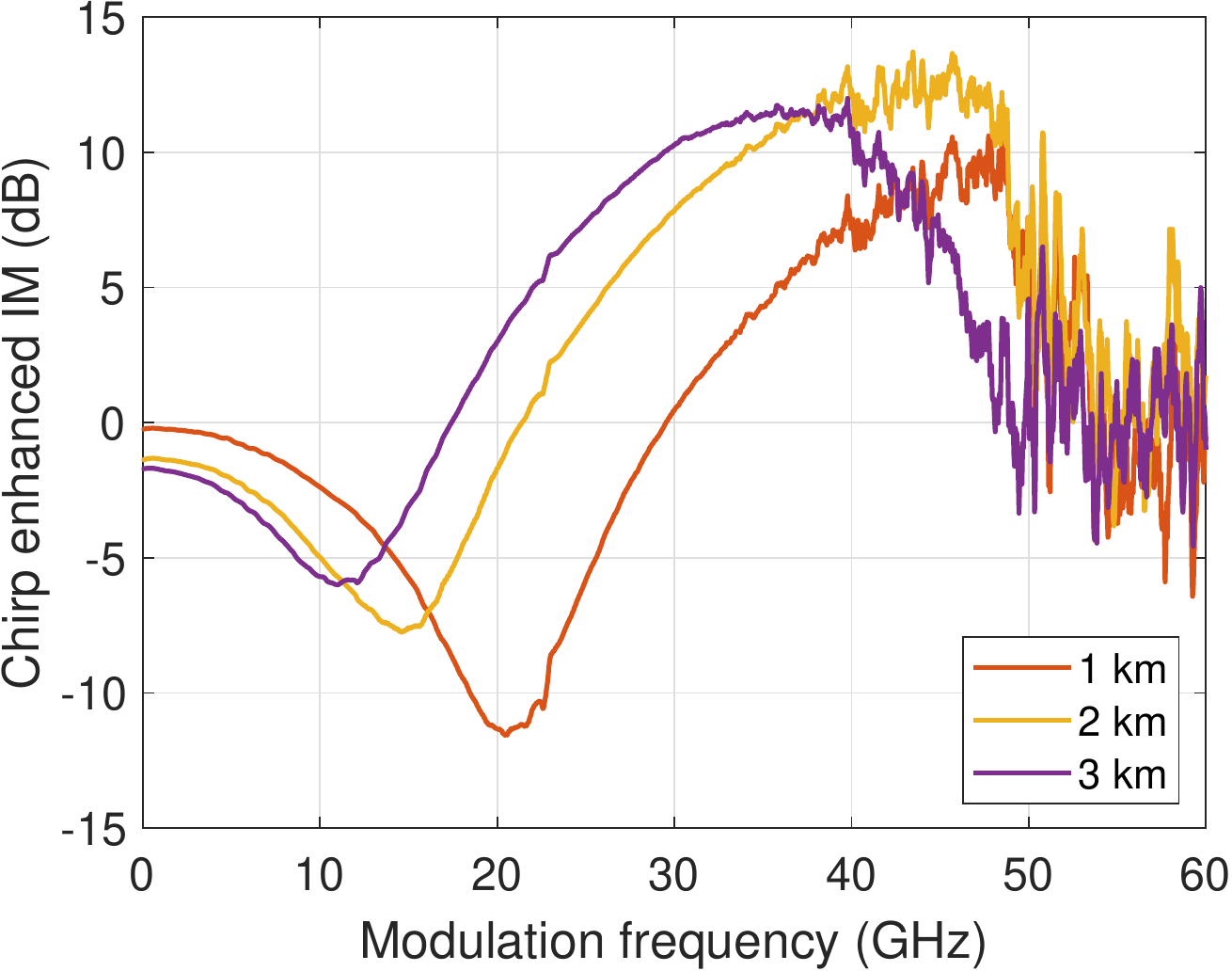}}
\caption{The S21 parameters of the VCSEL link with various SSMF lengths normalized to optical B2B transmission to isolate the effect of the combination of intensity modulation, phase modulation and chromatic dispersion in the fiber. }
\label{fig14}
\end{figure}

Moreover, the attenuations of the dispersion notch (-40dB @ 27~GHz in Fig. \ref{fig13}), are much stronger than those in the 1~km MCF case (-25dB @ 23~GHz in Fig. \ref{fig9}). This effect is due to the interplay among intensity and phase modulation in the VCSEL, and the chromatic dispersion of the fiber. To obtain more insight into this effect, the S21-parameters of various fiber lengths are measured and normalized to that in optical B2B transmission, as shown in Fig. \ref{fig14} \cite{{Kerrebrouck2}}. It is evident that the notch becomes deeper with increasing fiber lengths.

\begin{figure}[!t]
\centerline{\includegraphics[width=\columnwidth]{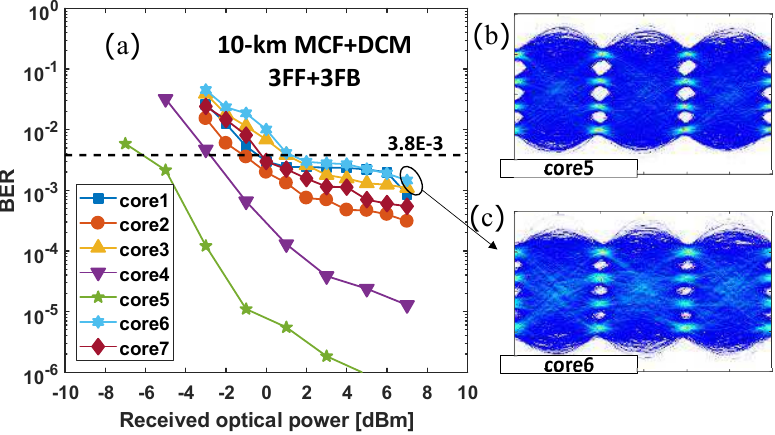}}
\caption{BER results for 50 GBd PAM-4 signal after 10-km MCF and eye diagrams at 7 dBm RoP for the best and worst cores.}
\label{fig15}
\end{figure}

 3-tap FF and 3-tap FB equalizers are proven to be sufficient at 50 Gbaud. Additionally, performance difference among the cores is also observable in the bit error rate measurement and selected eye diagrams are shown in Fig. \ref{fig15} (b) and (c). As mentioned before, these deviations are in accordance with the differences in the characterized frequency responses. Specific amplitude-level dependent symbol decision methods or nonlinear equalization techniques can be incorporated in the future work to mitigate the eye skew, leading to potentially better BER performance and/or higher baud rates. More pronounced differences are observed for core 5 in the 10-km MCF link, as shown in Fig. \ref{fig13}. We attribute this to the filtering effect induced by the fan-in/fan-out devices. 

\subsection{56~Gbaud 1-km}
\begin{figure}[!t]
\centerline{\includegraphics[width=0.9\columnwidth]{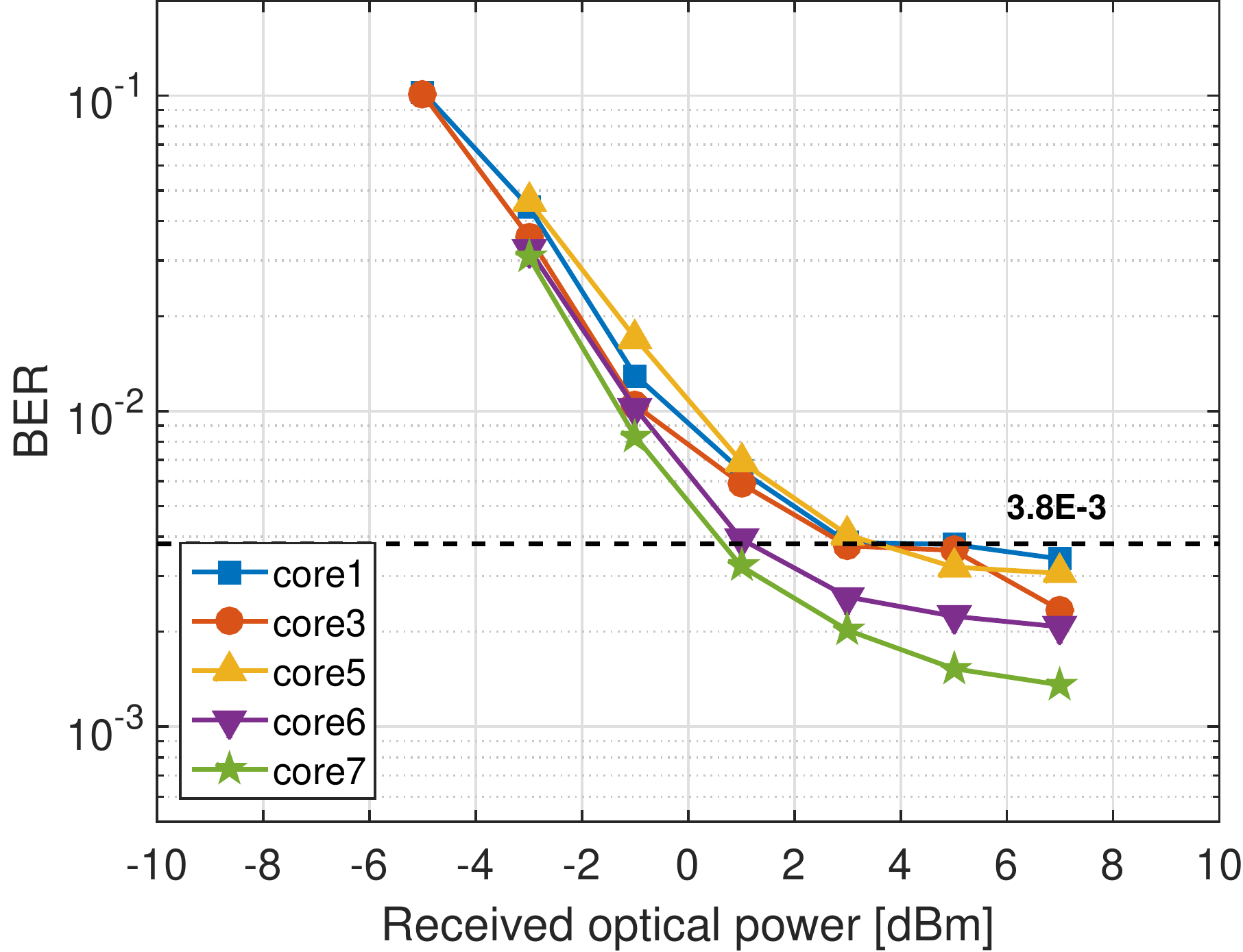}}
\caption{BER results for 56 Gbaud PAM-4 signal after 1-km MCF.}
\label{fig16}
\end{figure}

To explore the limit of the link, 56~Gbaud 4-PAM is transmitted over 1-km fiber without DCM. A 7-taps symbol-spaced FFE is used together with a 7-taps DFE. Only core 1, 3, 5, 6 and 7 reach a BER below the HD-FEC 7\% limit of 3.8 $\times$ 10$^{-3}$, as shown in Fig. \ref{fig16}. 

\section{Conclusion}
We have experimentally demonstrated high-speed short-reach optical transmission with long-wavelength SM-VCSEL over MCF. Up to 70 Gbaud PAM-4 has been transmitted with a VCSEL and MCF in B2B. We have successfully achieved up to 50~Gbaud PAM-4 transmission over 1-km dispersion-uncompensated and over 10-km dispersion compensated transmission through all the cores of the MCF links. Pre-equalization based on end-to-end link characterization in combination with simple offline post-equalization techniques has been proven to be sufficient for such optical links. 
VCSEL performance is extremely prospective for applications in short-reach optical interconnects. Further improvements can be obtained by further decreasing the chirp effect in the 1550~nm VCSEL or paying attention to the interplay between the dispersion properties of the link and the chirp of the VCSEL, to unlock higher baud rates  and/or achieve longer fiber length.

\section*{Acknowledgment}
The authors would like to thank the successful collaboration between Ghent University, Belgium and KTH Royal Institute of Technology, Sweden. This work was partly supported by the Natural Science Foundation of China (61331010, 61722108, 61775137, 61671212, and 61550110240),H2020 MCSA-IF Project NEWMAN (grant 752826), H2020 5GPPP 5G-PHOS research program (grant 761989), European Commission through the FP7 project MIRAGE (ref.318228), H2020 project WIPE (ref.688572), Swedish Research Council (VR), the Swedish Foundation for Strategic Research (SSF), G\"oran Gustafsson Foundation, Swedish ICT-TNG, the Celtic-Plus sub-project C2015/3-5 EXTEND \& SENDATE FICUS funded by Vinnova.


\begin{thebibliography}{00}

\bibitem{Cisco} Cisco, ``global cloud index: forecast and methodology, 2016-2021,'' [Online].
Available: http://www.cisco.com/

\bibitem{datacenter} Infinera, ``Rise of High-Capacity Data Center Interconnect in Hyper-Scale Service Provider Systems,'' [Online].
Available: https://www.infinera.com/

\bibitem{Kuchta} D. M. Kuchta, ``High capacity VCSEL-based links,'' presented at the Optical Fiber Communication Conf., Los Angeles, CA, USA, 2017, Paper Tu3C.4.

\bibitem{Richardson} D. J. Richardson, J. M. Fini, and L. E. Nelson, ``Space-division multiplexing in optical fibres,'' \emph{Nature Photon.}, vol. 7, no. 5, pp. 354-362, Apr. 2013.

\bibitem{Karppinen} M. Karppinen, A. Tanskanen, V. Heikkinen, P. My\"oh\"anen, N. Salminen, J. Ollila, O. Tapaninen, P. Westbergh, J. Gustavsson, A. Larsson, R. Safaisini, R. King, M. Ko, D. Kissinger, A. \c{C}. Ulusoy, T. Taunay, L. Bansal, L. Gr\"uner-Nielsen, E. Kehayas, J. Edmunds and L. Stampoulidis ``Integration of 150 Gbps/fiber optical engines based on multicore fibers and 6-channel VCSELs and PDs,'' in \emph{Proc. SPIE}, Vol. 9753, \emph{Optoelectronic Interconnects} XVI, (2016).

\bibitem{Watanabe} K. Watanabe, T. Saito, K. Kawasaki, M. Iwaya, T. Ando, K. Suematsu, and M. Shiino, ``MPO type 8-multicore fiber connector with physical contact connection,'' \emph{J. Light. Technol.}, vol. 34, no. 2, pp. 351-357, Jan. 2016.

\bibitem{Kerrebrouck1} J. V. Kerrebrouck, L. Zhang, R. Lin, X. Pang, A. Udalcovs, O. Ozolins, S. Spiga, M. C. Amann, G. Van Steenberge, L. Gan, M. Tang, S. Fu, R. Schatz, S. Popov, D. Liu, W. Tong, S. Xiao, G. Torfs, J. Chen, J. Bauwelinck and X. Yin, ``726.7-Gb/s 1.5-$\mu$m single-mode VCSEL discrete multi-tone transmission over 2.5-km multicore fiber,'' presented at the Optical Fiber Communication Conf., Los Angeles, CA, USA, 2017, Paper M1I.2.

\bibitem{Eiselt} N. Eiselt, H. Griesser, J. Wei, A. Dochhan, R. Hohenleitner, M. Ortsiefer, M. Eiselt, C. Neumeyr, J. J. V. Olmos and I. T. Monroy, ``Experimental demonstration of 56 Gbit/s PAM-4 over 15 km and 84 Gbit/s PAM-4 over 1 km SSMF at 1525 nm using a 25G VCSEL,'' presened a the European Conf. Exhibition Optical Communciation, D\"{u}sseldorf, Germany, 2016, Paper Th.1.C.1.

\bibitem{Xie} C. Xie, P. Dong, S. Randel, D. Pilori, P. Winzer, S. Spiga, B. K\"ogel, C. Neumeyr and M. C. Amann, ``Single-VCSEL 100-Gb/s short-reach system using discrete multi-tone modulation and direct detection,'' presented at the Optical Fiber Communication Conf., Los Angeles, CA, USA, 2015, Paper Tu2H.2. 

\bibitem{Pang} X. Pang, J. V. Kerrebrouck, O. Ozolins, R. Lin, A. Udalcovs, L. Zhang, S. Spiga, M. C. Amann, G. Van Steenberge, L. Gan, M. Tang, S. Fu, R. Schatz, G. Jacobsen, S. Popov, D. Liu, W. Tong, G. Torfs, J. Bauwelinck, X. Yin, and J. Chen, ``7$\times$100 Gbps PAM-4 Transmission over 1-km and 10-km 
Single Mode 7-core Fiber using 1.5-$\mu$m SM-VCSEL,'' presented at the Optical Fiber Communication Conf., San Diego, CA, USA, 2018, Paper M1I.4.

\bibitem{Spiga1} S. Spiga, W. Soenen, A. Andrejew, D. M. Schoke, X. Yin, J. Bauwelinck, G. Boehm, and M. C. Amann, ``Single-mode high-speed 1.5-$\mu$m VCSELs,'' \emph{J. Light. Technol.}, vol. 35, no. 4, pp. 727-733, Aug. 2017.

\bibitem{Spiga2} S. Spiga, D. Schoke, A. Andrejew, M. M\"uller, G. Boehm, and M. C. Amann, ``Single-mode 1.5-$\mu$m VCSELs with 22-GHz small-signal bandwidth,'' presented at the Optical Fiber Communication Conf., Anaheim, CA, USA, 2016, Paper Tu3D.4.

\bibitem{YOFC} YOFC multi-core fiber MCF-7-42/150/250 product datahseet. [Online]. Available: http://en.yofc.com/wcs/Upload/201704/58ef2fee6b6c7.pdf

\bibitem{Li} B. Li, Z. Feng, M. Tang, Z. Xu, S. Fu, Q. Wu, L. Deng, W. Tong, S. Liu, and P. P. Shum, ``Experimental demonstration of large capacity WSDM optical access network with multicore fibers and advanced modulation formats,'' \emph{Opt. Express.}, vol. 23, no. 9, pp. 10997-11006, Feb. 2015.

\bibitem{Tu} J. Tu, K. Saitoh, M. Koshiba, K. Takenaga, and S. Matsuo, ``Design and analysis of large-effective-area heterogeneous trench-assisted multi-core fiber,'' \emph{Opt. Express.}, vol. 20, no. 14, pp. 15157-15170, Jun. 2012.

\bibitem{Castro} J. M. Castro, R. J. Pimpinella, B. Kose, Y. Huang, A. Novick and B. Lane, ``Eye skew modeling, measurements and mitigation methods for VCSEL PAM-4 channels at data rates over 66 Gb/s,'' presented at the Optical Fiber Communication Conf., Los Angeles, CA, USA, 2017, Paper W3G.3. 

\bibitem{Kerrebrouck2} J. V. Kerrebrouck, H. Li, S. Spiga, M. C. Amann, X. Yin, J. Bauwelinck, P. Demeester and G. Torfs, ``10 Gb/s radio-over-fiber at 28 GHz carrier frequency link based on 1550 nm VCSEL chirp enhanced intensity modulation after 2 km Fiber,'' presented at the Optical Fiber Communication Conf., San Diego, CA, USA, 2018, Paper W1F.1.

\bibitem{Szczerba} K. Szczerba, P. Westbergh, M. Karlsson, P. A. Andrekson, and A. Larsson, ``70 Gbps 4-PAM and 56 Gbps 8-PAM using an 850 nm VCSEL,'' \emph{J. Light. Technol.}, vol. 33, no. 7, pp. 1395-1401, Apr. 2015.

\bibitem{pam4_designcon} H. Zhang, B. Jiao, Y. Liao, G. Zhang,``PAM4 signaling for 56G serial link applications a tutorial,''  presented at DesignCon., Santa Clara, CA, USA, 2016.

\end{thebibliography}
\end{document}